\documentclass[12pt]{article}
\usepackage{amsmath,amssymb}

\newcommand{\bea}{\begin{eqnarray}}
\newcommand{\eea}{\end{eqnarray}}
\newcommand{\be}{\begin{equation}}
\newcommand{\ee}{\end{equation}}
\newcommand{\vs}[1]{\vspace{#1 mm}}

\newcommand{\dsl}{\pa \kern-0.5em /}

\newcommand{\pa}{\partial}

\newcommand{\nn}{\nonumber\\}

\begin{document}
\topmargin 0mm
\oddsidemargin 0mm

\begin{flushright}

USTC-ICTS/PCFT-23-21\\

\end{flushright}

\vspace{2mm}

\begin{center}

{\Large \bf Understanding the open string pair production\\
 of the Dp/D0 system}

\vs{10}

{\large J. X. Lu}

\vspace{4mm}

{\em
Interdisciplinary Center for Theoretical Study\\
 University of Science and Technology of China, Hefei, Anhui
 230026, China\\
 \medskip
 Peng Huanwu Center for Fundamental Theory, Hefei, Anhui 230026, China\\ 
}

\end{center}

\vs{10}

\begin{abstract}
 Consider a system consisting of D$p$ and D$p'$, placed parallel at a separation and with $p - p' = 2 n$ (assuming $p \ge p'$, the integer $n \ge 0$ and $p' > 0$). When either D$p$ or D$p'$ carries a worldvolume electric flux, one in general expects a non-vanishing open string pair production due to the pair of virtual open string/anti open string connecting the two D branes under the action of the applied flux.  However, this will not be true  for $p' = 0$ and $p = 2, 4, 6$ when the D$p$ carries a pure electric flux. In this note, we will explore the case for which a finite non-vanishing open string pair production rate  can indeed be produced when  a certain worldvolume  flux is applied to the Dp brane and understand the physics behind.  
    \\
\end{abstract}

\newpage

\section{Introduction}
A D-sting or a D1 brane can be combined with a fundamental string or F-string to form the so-called 1/2 BPS non-threshold bound state (F, D1) \cite{Schwarz:1995dk}.  This had been generalized to give a general 1/2 BPS non-threshold (F, Dp) for $1 \le p \le 6$ via T-dualities in \cite{Lu:1999qia, Lu:1999uca}\footnote{The existence of such bound states  was shown in \cite{Witten:1995im}. Related works using mixed boundary conditions for these bound states were discussed in \cite{Arfaei:1997hb, Sheikh-Jabbari:1997qke}.}. The corresponding D-brane boundary state carrying a constant worldvolume electric flux was given later in \cite{DiVecchia:1999uf}.   

For an isolated Dp brane in type II superstring theories, the quantum fluctuations give rise to virtual oriented open strings, each of which has its two ends attaching to the D brane and carrying charges equal in amount but opposite in sign.  For $p > 0$, unlike in QED, applying a constant worldvolume  electric flux to this D-brane does not give rise to a pair production of the charge/anti-charge carried by the two ends of the virtual open string. This is because the force acting on either end of the virtual open string due to the constant electric flux is equal in magnitude but opposite in direction. As such the electric forces can only stretch the open string but cannot break it unless the electric flux  reaches its critical value for which the whole D brane system becomes unstable. This is consistent with the fact mentioned above that the (F, Dp) is a stable 1/2 BPS non-threshold bound state when the electric flux is below its critical value.  In other words, an isolated Dp brane cannot give rise to the open string pair production  when a non-critical electric flux is applied.   

The simplest system for producing such a pair production is one consisting of two D branes with the same or different spatial dimensionality, placed parallel at a separation, when an electric flux is applied on either of the branes.  This has been studied recently for a system of D$p$/D$p'$ with $p - p' = 2 n$ ($n = 0, 1, 2, 3$ and $p' > 0$) in type II superstring theories in a series of publications by the present author and his collaborators \cite{Lu:2017tnm, Lu:2018suj, Lu:2018nsc, Jia:2018mlr, Jia:2019hbr, Lu:2019ynq, Lu:2020hml}  in the spirit of Schwinger pair production in QED \cite{Schwinger:1951nm}. The open string pairs produced are those connecting the two D branes and are along the directions transverse to both D branes, therefore along the extra directions with respect to the brane observers.  In other words, this pair production can provide a means to detect the existence of extra dimensions if it can be detected and if the brane-world picture is employed \cite{Lu:2018nsc, Lu:2019ynq, Lu:2020hml}.

However, this is not quite true when $p' = 0$.  The reason is simple. Unlike the D$p$/D$p'$ with $p \ge p' > 0$ mentioned above, the two ends of the virtual open string/anti-open string attached on the D0 cannot be separated in general  while this is not an issue for the case when $p' > 0$ since the two ends can be separated along the brane spatial direction(s) (for the former, we don't have such 
a brane spatial direction to separate the two ends unless we break the virtual strings).  

The purpose of the present note is to explore the possibility that there exists a finite non-vanishing open string pair production rate for the Dp/D0 for $p = 2, 4, 6$, respectively, when the fluxes on the Dp are chosen properly while keeping both Dp and D0 largely stable.  We will also provide the physics behind such a non-vanishing pair production. 

The general properties for having such a non-vanishing pair production are basically the same for all the D$p$/D0 systems with $p = 2, 4, 6$.  However, there exist differences for each of the three cases considered. We will give each of these cases a detail consideration.  

For such a system,  the Dp worldvolume Lorentz symmetry is no longer preserved due to the D0 brane but the SO(p) rotation one is still there.  So we expect that the interaction amplitude will respect this SO(p) rotation symmetry. For this reason, we can choose the electric flux on Dp along the $`1'$ direction for simplicity and without loss of generality. The magnetic flux on Dp can be chosen more generally when $p > 2$ but for the purpose of this paper, we choose the same one, which is the most general one for the $p = 2$ case, for all the three cases.  This choice actually captures the essential feature needed. In other words, we choose the fluxes $\hat F_{p}$ on Dp for $p = 2, 4, 6$, respectively, as 
\be\label{dpflux}
 \hat F_{p} =\left( \begin{array}{ccccc}
0 & \hat f & 0&0 & \ldots\\
- \hat f&0&\hat g&0&\ldots\\
0&- \hat g&0&0&\ldots\\
0&0&0&0&\ldots\\
 \vdots&\vdots&\vdots&\vdots&\ddots
 \end{array}\right)_{(1 + p)\times (1 + p)},
\ee
where the dimensionless flux $\hat F_{\alpha\beta} = 2 \pi \alpha' F_{\alpha\beta}$ with $F_{\alpha\beta}$ the usual electromagnetic flux,  $2 \pi \alpha'$ the inverse of fundamental string tension and $\alpha, \beta = 0, 1, \cdots, p$. Here the dimensionless electric flux $ 0 \le |\hat f| \le \sqrt{1 + {\hat g}^{2}}$ with the up limit determined by the underlying Born-Infeld factor and the dimensionless magnetic flux  $|\hat g| \ge 0$.  Note that  $|\hat f_{\rm critical}| = \sqrt{1 + {\hat g}^{2}}$ corresponds to the critical electric flux for the Dp for which the Dp becomes unstable via the cascade open string pair production. When $|\hat g| = 0$, the critical electric flux $|\hat f_{\rm critical}| = 1$, the familiar one. 

As will be demonstrated later, when $\hat g =0$, there is no chance to give rise to a finite non-vanishing open string pair production for the D$p$/D0 system. In other words,  we must keep 
$\hat g $ non-vanishing on Dp to have such a pair production rate.  As we will see, unlike the system of D$p$/D$p'$ with $p \ge p' > 0$,  the mere presence of an electric flux ${\hat f}$ on D$p$ is not enough to give rise to a non-vanishing open string production and we further need $|{\hat f}| > 1$ which is only possible if a non-vanishing $\hat g$ is present.  This applied electric flux actually breaks the virtual open string and anti open string connecting the D$p$ and D0 to give rise to the open string pair production. 

This paper is organized as follows.  In section 2, we focus on the system of D2/D0 and analyze the stringy interaction between the D2 and D0 to reveal the condition for which a non-vanishing open string pair production can arise. Depending on the value of ${\hat f}$, we have three subcases to consider and the open string pair production occurs only for $1 < {\hat f}^{2} < 1 + {\hat g}^{2}$.  However,  understanding each of these subcases directly finds to be difficult. Instead, we achieve them much easier via  a particular system consisting of two D2, placed parallel at the same separation as the D2 and D0, with one D2 carrying certain fluxes and the other carrying no flux.  We will discuss each of three subcases in detail. In section 3, we move to discuss the D4/D0 system with D4 carrying the flux as given in (\ref{dpflux}) for $p = 4$. The discussion goes along the same line as that for the D2/D0 given in Section 2 but the detail is different. For example, the interaction nature is different and the system used to understand the D4/D0 is  D4/D2 for the present case. In section 4, we discuss the D6/D0 with the D6 carrying the flux as given in (\ref{dpflux}) for $p = 6$. Again, the discussion goes also the same line as that for D2/D0 but the detail is different, too.  The system used to understand D6/D0 is now D6/D2. We will spell the detail analysis in this section. We conclude this paper in section 5.  

\section{The D2/D0 case}

For this, we use the D-brane boundary state representation \cite{Di Vecchia:1999fx} and  follow the trick used in \cite{Jia:2019hbr} to compute the closed string tree-level cylinder interaction amplitude between the D2 and D0. Then we use the Jacobi transformation to obtain the corresponding open string one-loop annulus amplitude as
\bea\label{20annulusA}
\Gamma_{2, 0} &=& \frac{2 V_{1} \sqrt{1 - {\hat f}^{2}}}{(8 \pi^{2} \alpha')^{1/2}} \int_{0}^{\infty} \frac{d t}{t^{3/2}} e^{- \frac{y^{2} t}{2\pi \alpha' }} \frac{\left(1 - \cosh \pi \nu t\right)^{2}}{\sinh \pi \nu t} \nn
&\,& \times \prod_{n = 1}^{\infty} \frac{\left(1 - 2 |z|^{2n} \cosh \pi \nu t + |z|^{4n}\right)^{4}}{\left(1 - |z|^{2n}\right)^{6} \left(1 - 2 |z|^{2n} \cosh 2 \pi \nu t + |z|^{4n}\right)},
\eea
where   $|z| = e^{- \pi t}$, $y$ is the brane separation between the D2 and D0, $V_{1}$ is the volume of D0 worldvolume, i.e. $V_{1} = 2 \pi \delta(0)$,   and the parameter $\nu$ is given by
\be\label{parameternu}
\tan \pi \nu = \frac{\sqrt{1 - {\hat f}^{2}}}{|\hat g|}
\ee
with $0 <  {\hat f}^{2} < 1 + {\hat g}^{2}$.  We have three distinct subcases to study: 1) $0 < {\hat f}^{2} < 1$, 2) ${\hat f}^{2} = 1$  and 3) $1 < {\hat f}^{2} < 1 + {\hat g}^{2}$.\\
\noindent
{\bf Subcase 1):  $0 < {\hat f}^{2} < 1$.}\\
For this subcase,  the parameter $\nu$ is real and $\nu \in (0, 1/2)$ from (\ref{parameternu}). It is clear that the amplitude  (\ref{20annulusA}) is positive for this subcase  and therefore the interaction between the D2 and D0 is attractive according to our conventions. The integrand of the amplitude has no singularities, in particular no simple poles, for any finite $t$ (note that $t > 0$), indicating that no open string pair production arises, even though there is an electric flux present.  This must indicate that the applied electric flux does not break the virtual open string/anti open string pair connecting the D2 and D0 when ${\hat f}^{2} < 1$ since this is the only possibility to give rise to the open string pair production for the system under consideration. One would expect a non-vanishing open string pair production when ${\hat f}^{2} > 1$ since to this pair the critical electric flux is expected to be ${\hat f}^{2} = 1$. The integrand of the amplitude does have a potential singular behavior for $t \to \infty$. Let us take a close look at it.  For large $t$, the potential singular behavior of the integrand is controlled by 
\be
\sim \frac{e^{- \frac{y^{2} t}{2 \pi \alpha'}}}{t^{3/2}}  \frac{(1 - \cosh \pi \nu t)^{2}}{\sinh \pi \nu t} \sim \frac{e^{- \frac{t}{2\pi \alpha'} \left(y^{2} - 2 \pi^{2} \nu \alpha'\right)}}{t^{3/2}},
\ee
which vanishes when $y \ge \pi \sqrt{2 \nu \alpha'}$ and blows up when $y < \pi \sqrt{2 \nu \alpha'}$, indicating a tachyonic instability for the latter.  Note that $|z| \to 0$ and each factor in the infinite product in the integrand goes to unity for large $t$. So all the properties for this case are almost identical to, for example, a D2/D2 system carrying only worldvolume magnetic flux which we will come to discuss later on. \\
\noindent
{\bf Subcase 2: ${\hat f}^{2} = 1, |{\hat g}| \neq 0$.}\\ 
For this, from (\ref{parameternu}), we have $\nu = 0$. Note that now each factor in the infinite product in the integrand becomes unity.   It is not difficult to check that the amplitude  (\ref{20annulusA}) now vanishes. We will  come to explain this result later on.\\
\noindent
{\bf Subcase 3: $1 < {\hat f}^{2} < 1 + {\hat g}^{2}$.}\\
For this case, the parameter $\nu$ given in (\ref{parameternu}) becomes imaginary and if we set $\nu = i \nu_{0}$, we have 
\be\label{parameternu0}
\tanh \pi \nu_{0} = \frac{\sqrt{{\hat f}^{2} - 1}}{|\hat g|},
\ee
with $\nu_{0} \in (0, \infty)$.
The amplitude (\ref{20annulusA}) now becomes
\bea\label{20annulusAmp}
\Gamma_{2, 0} &=& \frac{2 V_{1} \sqrt{{\hat f}^{2} - 1}}{(8 \pi^{2} \alpha')^{1/2}} \int_{0}^{\infty} \frac{d t}{t^{3/2}} e^{- \frac{y^{2} t}{2\pi \alpha' }} \frac{\left(1 - \cos \pi \nu_{0} t\right)^{2}}{\sin \pi \nu_{0} t}\nn
&\,&  \times \prod_{n = 1}^{\infty} \frac{\left(1 - 2 |z|^{2n} \cos \pi \nu_{0} t + |z|^{4n}\right)^{4}}{\left(1 - |z|^{2n}\right)^{6} \left(1 - 2 |z|^{2n} \cos 2 \pi \nu_{0} t + |z|^{4n}\right)}.
\eea
The integrand of amplitude (\ref{20annulusAmp}) has simple poles occurring at $\sin\pi \nu_{0} t = 0, \cos\pi \nu_{0} t \neq 1$, indicating the decay of this system via the open string pair production. These poles are at  
\be
\pi \nu_{0} t_{k} = (2 k - 1) \pi \to t_{k} = \frac{2 k - 1}{\nu_{0}}, \quad k = 1, 2, \cdots .
\ee
The decay rate can be computed as the sum of the residues of the integrand in (\ref{20annulusAmp}) at these poles times $\pi$ per unit D0 worldvolume following  \cite{Bachas:1992bh} as
\bea\label{decay-rate20}
{\cal W}_{2, 0}  &=& - \frac{2 {\rm Im} \Gamma_{2,0}}{V_{1}}\nn
&=& \frac{2^{4} \sqrt{{\hat f}^{2} - 1}}{\nu_{0} (8 \pi^{2} \alpha')^{1/2}}  \sum_{k = 1}^{\infty} \left(\frac{\nu_{0}}{2 k - 1}\right)^{3/2} \, e^{- \frac{(2 k - 1) y^{2}}{2\pi \nu_{0} \alpha'}} \prod_{n = 1}^{\infty}
 \frac{\left(1 + |z_{k}|^{2n}\right)^{8}}{\left(1 - |z_{k}|^{2n}\right)^{8}},
\eea
with $|z_{k}| = e^{- \frac{(2 k - 1)\pi}{\nu_{0}}}$. The open string pair production rate\footnote{There existed confusions and still remain so in some papers about the decay rate or pair production probability per unit volume and time (say, the present ${\cal W}_{2,0}$ (\ref{decay-rate20})) with the pair production rate (say, the present ${\cal W}^{(1)}_{2,0}$ given below (\ref{pairp-rate20})). However, these two quantities are different. We give a rather detail discussion of this difference in the Appendix.}
    is then given as the $k = 1$ term of the above decay rate following \cite{nikishov} as
\be\label{pairp-rate20}
{\cal W}^{(1)}_{2, 0} = \frac{2^{4} \sqrt{{\hat f}^{2} - 1}}{(8 \pi^{2} \alpha')^{1/2}}  \nu_{0}^{1/2} \, e^{- \frac{ y^{2}}{2\pi \nu_{0} \alpha'}} \prod_{n = 1}^{\infty}
 \frac{\left(1 + |z_{1}|^{2n}\right)^{8}}{\left(1 - |z_{1}|^{2n}\right)^{8}},
 \ee
 with $|z_{1}| = e^{- \pi/\nu_{0}}$. So indeed we have a finite non-vanishing open string pair production rate for the present subcase. 
 
 At a first look, it would be surprised to have this since one doesn't in general expect that the two ends of the virtual open string/anti-open string attached on the D0 brane can be separated away.  But with this finite non-vanishing rate, we must then have that the applied electric flux $\hat f$ falling in the range of $1 < {\hat f}^{2} < 1 + {\hat g}^{2}$  either breaks the virtual open string pair to make it  become real or pull the two ends of the virtual pair attached on the D0 away from it to give rise to a new pair of D0/anti D0 with the end of one of the open strings in the pair attached on, for example, the D0 in the pair of D0/anti D0 while the other end attached on the anti D0.  The end result for the former is actually the same as for the latter.   In other words, for either scenario,  the end result is the same, i.e., we have an open string pair whose two ends on the D2 are pulled away along the D2 by the applied electric flux while the other two ends, one is attached on a newly created D0 while the other is on the newly created anti D0, are also pulled away by this flux.
 
 Each of the above three subcases is determined by the corresponding range of the applied electric flux $|{\hat f}|$ and the applied magnetic flux ${\hat g}$ does not seem to play much role except that its appearance allows $|{\hat f}| > 1$ so that there is a possibility for the open string pair production as demonstrated above. The underlying physics behind does not seem to be clear or intuitive for each of the three distinct subcases. In other words, we don't have a clear picture why when $0 < {\hat f}^{2} < 1$, the interaction is attractive, ${\hat f}^{2} = 1$ the interaction vanishes and $1 < {\hat f}^{2} < 1 + {\hat g}^{2}$, there exists  the open string pair production.  
 
 While we don't have a direct understanding of the three subcases, we  come with  an alternative or possibly a better explanation to each of these subcases.  It is known that a magnetic flux on a Dp brane stands for the presence of a co-dimension 2 D-branes inside the original Dp brane, forming the so-called 1/2 BPS non-threshold bound state (D(p - 2), Dp) \cite{Witten:1995im, Breckenridge:1996tt, Costa:1996zd, Gava:1997jt, Di Vecchia:1997pr, Sheikh-Jabbari:1997qke}.  The present author and his collaborator also noticed in \cite{Lu:2009pe} that the co-dimension 2 brane relative to the original brane when considering their interaction behaves effectively as a magnetic flux.  With all these in mind and for simplicity, we consider a system of two D2, placed parallel at separation with one of them carrying the general flux as in (\ref{dpflux}) for $p = 2$ but here denote it differently as
 \be\label{d2flux-new}
\hat F = \left(\begin{array}{ccc}
0 &  f & 0\\
-  f & 0 &  g\\
0 & - g & 0 \end{array} \right),
\ee 
(here either $f$ or $g$ without a hat)  while the other D2 carrying no flux.  Then following \cite{Lu:2018suj, Jia:2019hbr},  we compute the open string one-loop annulus interaction between these two D2 branes as
 \bea\label{22annulusAmp}
 \Gamma_{2, 2} &=& \frac{2^{2} V_{3} \sqrt{ g^{2} - f^{2}}}{(8 \pi^{2} \alpha')^{3/2}} \int_{0}^{\infty} \frac{d t}{t^{3/2}} \, e^{- \frac{y^{2} t}{2\pi \alpha'}} \frac{(1 - \cosh \pi \bar\nu t)^{2}}{\sinh\pi \bar\nu t} \nn
 &\,& \times \prod_{n = 1}^{\infty} \frac{(1 - 2 |z|^{2n} \cosh \pi \bar\nu t + |z|^{4n})^{4}}{(1 - |z|^{2n})^{6} (1 - 2 |z|^{2n} \cosh 2 \pi \bar\nu t + |z|^{4n})},
 \eea
 where $|z| = e^{- \pi t}$, $y$ is the separation between the two D2 (here taken the same as that between the D2 and D0),  $V_{3}$ is the volume of the D2 worldvolume and the parameter $\bar\nu$ is given by
 \be\label{parameternup}
 \tan\pi \bar\nu = \sqrt{ g^{2} - f^{2}}.
 \ee 
 This one-loop annulus amplitude structurally looks the same as that given in (\ref{20annulusA}) for the D2/D0 system.  Apart from the trivial factor $ 2 V_{2} g /  (8 \pi^{2} \alpha')$, the two amplitudes are identical if we make the following identifications
 \be\label{parameterident}
 g = \frac{1}{\hat g}, \quad f = \frac{\hat f}{\hat g},
 \ee
 and with these we have also $\nu = \bar \nu$.  Note that for the present case, we need also to have $1 - f^{2} + g^{2} > 0$ to guarantee  the stability of the D2 carrying the flux (\ref{d2flux-new}). With the identifications of (\ref{parameterident}), $1 - f^{2} + g^{2} > 0$ gives $\left(1 - {\hat f}^{2}  + {\hat g}^{2}\right)/{\hat g}^{2} > 0$ which therefore implies $1 - {\hat f}^{2} + {\hat g}^{2} > 0$, a consistent result.
Therefore so long the interaction is concerned, with (\ref{parameterident}), the amplitude (\ref{22annulusAmp}) for the D2/D2 is essentially the same as that  given in (\ref{20annulusA}) for the D2/D0.  

We try now to use the amplitude (\ref{22annulusAmp}) to give a better understanding of the three subcases discussed above for the D2/D0 system.  Note that the amplitude  (\ref{22annulusAmp}) is a D2 worldvolume Lorentz invariant.  In other words, if we make a D2 worldvolume Lorentz transformation, the transformed system is physically equivalent to the original D2/D2 system so long the interaction is concerned.  With this, we in turn give an explanation to each of the three subcases discussed earlier for the D2/D0 system. 

Using the relations (\ref{parameterident}), it is clear that each of the three subcases considered for the D2/D0 has its correspondence to the present one for the D2/D2 as follows: $0 < {\hat f}^{2} < 1$ gives $g^{2} > f^{2}$,  ${\hat f}^{2} = 1$ gives $g^{2} =  f^{2}$ and $1 < {\hat f}^{2} < 1 + {\hat g}^{2}$ gives $g^{2} < f^{2}$. So we need to understand these three subcases for the D2/D2 system in what follows.

For $g^{2} > f^{2}$,  we can use the following Lorentz boost 
\be\label{g-Lorentz}
\Lambda_{\alpha}\,^{\beta} =\left(\begin{array}{ccc}
 \frac{g}{\sqrt{g^{2} -  f^{2}}}  &  0 &   \frac{f}{\sqrt{g^{2} -  f^{2}}} \\
0 & 1 &  0\\
 \frac{f}{\sqrt{g^{2} -  f^{2}}}  & 0 & \frac{g}{\sqrt{g^{2} -  f^{2}}} \end{array} \right),
\ee 
to transform the original flux (\ref{d2flux-new}) to a pure magnetic one  as
\be\label{transf}
\Lambda_{\alpha}\,^{\gamma} \Lambda_{\beta}\,^{\delta} \hat F_{\gamma\delta} \to \bar F_{\alpha\beta} =   \left(\begin{array}{ccc}
0 &  0 & 0\\
0 &0  &  \bar g \\
0 & - \bar g & 0 \end{array} \right),
\ee
where $\bar g = \sqrt{g^{2} - f^{2}}$, while the zero flux on the other D2 still remains so under this transformation.  Therefore so long the interaction is concerned, the original D2/D2 system is equivalent to the transformed D2/D2 with one D2 carrying a pure magnetic flux $\bar F$ as given above and the other D2 carrying no flux.  

For the transformed system, the interaction between the two D2 is known to be attractive and this can also be easily understood by the fact that there is no interaction between two parallel D2 when both of them carry no flux but there is an attractive interaction between a D2 carrying no flux and D0 branes which are represented here by the magnetic flux carried by one of the D2 given above. So the net interaction for the transformed system is attractive.  This then gives an explanation to the attractive interaction between the two D2 in the original D2/D2 system when $g^{2} > f^{2}$. This  in turn further gives an explanation to the attractive interaction for the D2/D0 system for the subcase of $0 < {\hat f}^{2} < 1$.

We now move to the second subcase $g^{2} = f^{2}$. To deal with this case properly, we take it as a limiting case of $f^{2} \to g^{2}$ or ${\hat f}^{2} \to 1$ from the previous one so that the Lorentz transformation of the above can still be used. In other words, we take a large Lorentz boost of (\ref{g-Lorentz}) in this limit (Note that $\det \Lambda = 1$ even in this limit). In this limit, 
$\bar g = 0$ and the transformed flux given in (\ref{transf}) vanishes.  In this limit, the interaction amplitude (\ref{22annulusAmp}) simply vanishes. This can be simply explained from the transformed system. 
After the transformation, the original D2/D2 system is transformed to a system of D2/D2 with neither of the D2's carrying any flux. So this is simply a system of two D2, placed parallel at a separation, with neither of them carrying any flux. We know that this system is still 1/2 BPS, breaking no SUSY, just like either of the D2's, and there is no net interaction between them.  

We come to the final subcase of $g^{2} < f^{2}$ with $1 - f^{2} + g^{2} > 0$, i.e. $1 < {\hat f}^{2} < 1 + {\hat g}^{2}$.  For this subcase, the parameter $\bar \nu = \nu = i \nu_{0}$ with $\nu_{0} \in (0, \infty)$, i.e. becoming imaginary and the amplitude (\ref{22annulusAmp}) is now
 \bea\label{22annulusA}
 \Gamma_{2, 2} &=& \frac{2^{2} V_{3} \sqrt{f^{2} - g^{2}}}{(8 \pi^{2} \alpha')^{3/2}} \int_{0}^{\infty} \frac{d t}{t^{3/2}} \, e^{- \frac{y^{2} t}{2\pi \alpha'}} \frac{(1 - \cos \pi \nu_{0} t)^{2}}{\sin\pi \nu_{0} t} \nn
 &\,& \times \prod_{n = 1}^{\infty} \frac{(1 - 2 |z|^{2n} \cos \pi \nu_{0} t + |z|^{4n})^{4}}{(1 - |z|^{2n})^{6} (1 - 2 |z|^{2n} \cos 2 \pi \nu_{0} t + |z|^{4n})},
 \eea
 where 
 \be\label{parameternup0}
 \tanh\pi \nu_{0} = \sqrt{ f^{2} - g^{2}}.
 \ee 
 The integrand of this amplitude is essentially the same as that given in (\ref{20annulusAmp}) for the D2/D0 case.  Therefore the discussion will go along the same line as what we did above for the D2/D0 case and will not repeat it here.  The decay rate of the present case can be computed as before and it is
 \bea\label{22decayr}
 {\cal W}_{2, 2} &=& - \frac{2 \,{\rm Im} \Gamma_{2, 2}}{ V_{3}}\nn
 &=& \frac{2^{5} \sqrt{f^{2} - g^{2}}}{\nu_{0} (8 \pi^{2} \alpha')^{3/2}} \sum_{k = 1}^{\infty} \left(\frac{\nu_{0}}{2 k - 1}\right)^{3/2}\, e^{- \frac{(2 k - 1) y^{2}}{2\pi \nu_{0}\alpha'}} \prod_{n = 1}^{\infty} \frac{\left(1 + |z_{k}|^{2n}\right)^{8}}{\left(1 - |z_{k}|^{2n}\right)^{8}},
 \eea
where $|z_{k}| = e^{- (2 k - 1)\pi/\nu_{0}}$.  The corresponding open string pair production rate, following \cite{nikishov}, is the $k = 1$ term of the above as
\be
{\cal W}^{(1)}_{2, 2} =  \frac{2^{5} \sqrt{f^{2} - g^{2}}}{(8 \pi^{2} \alpha')^{3/2}} \nu_{0}^{1/2} \, e^{- \frac{ y^{2}}{2\pi \nu_{0}\alpha'}} \prod_{n = 1}^{\infty} \frac{\left(1 + |z_{1}|^{2n}\right)^{8}}{\left(1 - |z_{1}|^{2n}\right)^{8}}, 
\ee
where $|z_{1}| = e^{- \pi/\nu_{0}}$.  As anticipated, here either rate differs from the corresponding one for D2/D0 by a trivial factor of $ 2 g/(8\pi^{2} \alpha')$.

We can use the following Lorentz boost 
\be\label{f-Lorentz}
\Lambda_{\alpha}\,^{\beta} =\left(\begin{array}{ccc}
 \frac{f}{\sqrt{f^{2} -  g^{2}}}  &  0 &   \frac{g}{\sqrt{f^{2} -  g^{2}}} \\
0 & 1 &  0\\
 \frac{g}{\sqrt{f^{2} -  g^{2}}}  & 0 & \frac{f}{\sqrt{f^{2} -  g^{2}}} \end{array} \right),
\ee  
to transform the original flux (\ref{d2flux-new}) to the following
\be\label{e-transf}
\Lambda_{\alpha}\,^{\gamma} \Lambda_{\beta}\,^{\delta} \hat F_{\gamma\delta} \to \bar F_{\alpha\beta} =  =   \left(\begin{array}{ccc}
0 &  \bar f & 0\\
- \bar f &0  &  0 \\
0 & 0 & 0 \end{array} \right),
\ee
where $\bar f = \sqrt{f^{2} - g^{2}}$.  In other words, for the transformed D2/D2 system, the D2 originally carrying no flux remains so after the transformation while the D2 originally carrying the flux (\ref{d2flux-new})  carries now a pure electric flux as given above after the transformation.  Given that for a system of two Dp branes with $p > 0$, placed parallel at a separation, when either of them carries an electric flux,  there is a non-vanishing open string pair production rate.   So the above clearly demonstrates that there is an open string pair production for the present subcase.  

Using the equivalence of the original D2/D2 and its transformed one in the sense described earlier, we clearly demonstrate and explain for the flux given in (\ref{d2flux-new}) the condition, i.e., $f^{2} > g^{2}$, with which the applied electric flux $f$ can give rise to an open string pair production.  For other subcases, this applied electric flux plays no role in producing the open string pair production and the underlying system is equivalent to one carrying pure magnetic one or carrying no flux at all.  This in turn helps us to understand the three  subcases for D2/D0 discussed earlier through the corresponding D2/D2 ones via the flux relations given in (\ref{parameterident}).  We would like to stress that even though we are certain about the open string pair production when $f^{2}/g^{2} > 1$ for the D2/D2 or equivalently $ {\hat f}^{2} > 1$ for the D2/D0, their respective actual productions  are different. For the former, it is a standard one, i.e., the applied electric flux pulls the virtual open string away from its anti one to make the virtual pair become a real one. While for the latter, the applied electric flux breaks each of the virtual open string and its anti one to make the virtual pair become a real one or pulls a pair of virtual D0 and its anti one out of the original D0 and make them real in the sense described earlier.  

Note that for the D2/D0, if we set the magnetic flux $\hat g = 0$ in (\ref{dpflux}) for $p = 2$, it is then impossible to have an open string pair production since we cannot have ${\hat f}^{2} > 1$ while maintaining $1 - f^{2}  > 0$ to have a stable D2 .  This can be easily understood using the D2/D2 system and the flux identifications given in  (\ref{parameterident}). Since if we set $\hat g  = 0$, this gives $g = 1/\hat g \to \infty$, we can no longer have $f^{2} > g^{2} \to \infty$ while maintaining $1 - f^{2} + g^{2} > 0$ such that the D2 carrying these fluxes remains  stable.

\section{The D4/D0 Case}

By the same token, we have also the same three subcases to consider for the D4/D0 system when the D4 carries the flux given in (\ref{dpflux}) for $p = 4$.  However, as will be clear, the detail for each of these subcases is different from the corresponding one for D2/D0.  Following \cite{Jia:2019hbr}, we have the open string one-loop annulus amplitude for the D4/D0 as
\be\label{40annulusA}
 \Gamma_{4, 0} = \frac{V_{1} \sqrt{1 - {\hat f}^{2}}}{(8 \pi^{2} \alpha')^{1/2}} \int_{0}^{\infty} \frac{d t}{t^{3/2}}\, e^{- \frac{y^{2} t}{2\pi \alpha'}} \frac{\left(\cosh \frac{\pi t}{2} - \cosh \pi \nu t\right)^{2}}{\sinh \frac{\pi t}{2} \sinh \pi \nu t} \prod_{n = 1}^{\infty} Z_{n},
 \ee
 where 
 \be \label{zn}
 Z_{n} = \frac{\left(1 - 2 |z|^{2n} \cosh\pi \left(\nu + \frac{1}{2}\right) t + |z|^{4n}\right)^{2} \left(1 - 2 |z|^{2n} \cosh \pi \left(\nu - \frac{1}{2}\right) t + |z|^{4n}\right)^{2}} {(1 - |z|^{2n})^{4} (1 - 2 |z|^{2n} \cosh \pi t + |z|^{4n})(1 - 2 |z|^{2n} \cosh 2\pi \nu t + |z|^{4n})}.
 \ee
In the above again $|z| = e^{ - \pi t}$,  $y$ is the separation between the D4 and D0, and the parameter $\nu$ continues to be given by (\ref{parameternu}).  We now discuss each of the three subcases in order.\\
 \noindent 
 {\bf Subcase 1: $0 < {\hat f}^{2} < 1$.}\\
 For this subcase,  the parameter $\nu$ is real and $\nu \in (0, 1/2)$  from (\ref{parameternu}).   The amplitude is clearly positive and therefore the interaction between the D4 and D0 is attractive. We will come to understand this later on using a different system. 
 
Note that the integrand of the amplitude has no singularities, in particular simple poles, for any finite $t > 0$. As for the D2/D0, this system has no open string pair production for this subcase.  This system has also a potential blowup for large $t$ since the integrand behaves as
 \be
 \sim \frac{e^{- \frac{y^{2} t}{2\pi \alpha'}}}{t^{3/2}} \, \frac{\left(\cosh \frac{\pi t}{2} - \cosh \pi \nu t\right)^{2}}{\sinh \frac{\pi t}{2} \sinh \pi \nu t} \to  \frac{e^{- \frac{t}{2\pi \alpha'} \left[y^{2} - 2 \pi^{2} \left(\frac{1}{2} - \nu \right) \alpha' \right]}}{t^{3/2}},
 \ee 
 which vanishes  when $y \ge \pi \sqrt{2 (1/2 - \nu)\alpha'}$ and blows up when $y <  \pi \sqrt{2 (1/2 - \nu)\alpha'}$ for large $t$ (Noticing here $\nu < 1/2$). The blowup case indicates a tachyonic instability of the system under consideration.  Again, all the properties of this subcase appear to 
 be almost identical to those of D4/D2 with the D4 carrying a pure magnetic flux while the D2 carrying no flux.  We will come to explain this subcase later on.\\
 \noindent 
 {\bf Subcase 2: ${\hat f}^{2} = 1$.}\\
 For this subcase, we have also $\nu = 0$ from (\ref{parameternu}).  However, unlike the D2/D0, the amplitude (\ref{40annulusA}) does not vanish and the resulting one can be obtained by taking the limits of ${\hat f} \to 1$ and $\nu \to 0$ from (\ref{40annulusA}).  Note that $\sinh\pi \nu t \to \pi \nu t, \cosh \pi \nu t \to 1$ and $\tan \pi \nu = \sqrt{1 - {\hat f}^{2}}/|\hat g| \to \pi \nu = \sqrt{1 - {\hat f}^{2}}/|\hat g|$ when we take ${\hat f}^{2} \to 1$ and $\nu \to 0$.  The resulting non-vanishing positive amplitude (therefore attractive interaction) is 
\be\label{40annulusA-new}
 \Gamma_{4, 0} = \frac{V_{1} |\hat g|}{(8 \pi^{2} \alpha')^{\frac{1}{2}}} \int_{0}^{\infty} \frac{d t}{t^{\frac{5}{2}}}\, e^{- \frac{y^{2} t}{2\pi \alpha'}} \frac{\left(\cosh \frac{\pi t}{2} - 1\right)^{2}}{\sinh \frac{\pi t}{2}} \prod_{n = 1}^{\infty} \frac{\left(1 - 2 |z|^{2n} \cosh \frac{\pi t}{2} + |z|^{4n}\right)^{4}}{\left(1 - |z|^{2n}\right)^{6} \left(1 - 2 |z|^{2n} \cosh \pi t + |z|^{4n}\right)}.
 \ee
We will come to understand this later on. \\
\noindent
{\bf Subcase 3: $1 < {\hat f}^{2} < 1 + {\hat g}^{2}$.} \\
For this subcase, the parameter $\nu$ again becomes imaginary  and if we set $\nu = i \nu_{0}$ with $\nu_{0} \in (0, \infty)$, the $\nu_{0}$ continues to satisfy (\ref{parameternu0}).  The amplitude (\ref{40annulusA}) becomes now
  \be\label{40annulusAmp}
 \Gamma_{4, 0} = \frac{V_{1} \sqrt{{\hat f}^{2} - 1}}{(8 \pi^{2} \alpha')^{1/2}} \int_{0}^{\infty} \frac{d t}{t^{3/2}}\, e^{- \frac{y^{2} t}{2\pi \alpha'}} \frac{\left(\cosh \frac{\pi t}{2} - \cos \pi \nu_{0} t\right)^{2}}{\sinh \frac{\pi t}{2} \sin \pi \nu_{0} t} \prod_{n = 1}^{\infty} Z_{n},
 \ee 
 where 
 \be \label{zn-i}
 Z_{n} = \frac{\left(1 - 2 |z|^{2n} \cosh\pi \left(i \nu_{0} + \frac{1}{2}\right) t + |z|^{4n}\right)^{2} \left(1 - 2 |z|^{2n} \cosh \pi \left(i \nu_{0} - \frac{1}{2}\right) t + |z|^{4n}\right)^{2}} {(1 - |z|^{2n})^{4} (1 - 2 |z|^{2n} \cosh \pi t + |z|^{4n})(1 - 2 |z|^{2n} \cos 2\pi \nu_{0} t + |z|^{4n})}.
 \ee 
 The simple poles of the above integrand occur at 
 \be
 \sin\pi \nu_{0} t = 0 \to \pi \nu_{0} t_{k} = k \pi \to t_{k} = \frac{k}{\nu_{0}}, \quad k = 1, 2, \cdots.
 \ee
 The decay rate of this system for this subcase can be computed to give
 \bea\label{40decayr}
 {\cal W}_{4, 0} &=& - \frac{2 {\rm Im} \Gamma_{4, 0}}{V_{1}}\nn
 &=& \frac{2 \sqrt{{\hat f}^{2} - 1}} {\nu_{0} (8 \pi^{2} \alpha')^{1/2}} \sum_{k = 1}^{\infty} ( - )^{k - 1} \left(\frac{\nu_{0}}{k}\right)^{3/2} e^{- \frac{k y^{2}}{2\pi \nu_{0} \alpha'}} \frac{\left(\cosh \frac{k \pi }{2\nu_{0}} - (-)^{k}\right)^{2}}{\sinh \frac{k \pi }{2 \nu_{0}}} \nn
 &\,& \times \prod_{n = 1}^{\infty} \frac{\left(1 - 2 (- )^{k}  |z_{k}|^{2n}  \cosh \frac{k \pi}{2 \nu_{0}} + |z_{k}|^{4n}\right)^{4}}{(1 - |z_{k}|^{2n})^{6} \left(1 - 2 |z_{k}|^{2n} \cosh\frac{k \pi}{ \nu_{0}} + |z_{k}|^{4n}\right)},
 \eea
 where $|z_{k}| = e^{- k \pi/\nu_{0}}$.  The open string pair production is the $k = 1$ term of the above and it is
 \be
 {\cal W}^{(1)}_{4, 0} = \frac{2 \nu_{0}^{1/2}\sqrt{{\hat f}^{2} - 1}} {(8 \pi^{2} \alpha')^{1/2}}  e^{- \frac{ y^{2}}{2\pi \nu_{0} \alpha'}} \frac{\left( 1 + \cosh \frac{\pi}{2\nu_{0}} \right)^{2}}{\sinh \frac{ \pi }{2 \nu_{0}}} \prod_{n = 1}^{\infty} \frac{\left(1 +  2   |z_{1}|^{2n}  \cosh \frac{\pi}{2 \nu_{0}} + |z_{1}|^{4n}\right)^{4}}{(1 - |z_{1}|^{2n})^{6} \left(1 - 2 |z_{1}|^{2n} \cosh\frac{\pi}{ \nu_{0}} + |z_{}|^{4n}\right)},
 \ee
 where $|z_{1}| = e^{- \pi/\nu_{0}}$.
 
Again it is hard to have a direct understanding of the above three subcases.  Alternatively, we use the D4/D2 system for this purpose.  For the D4/D2, we denote the flux on the D4 as
 \be\label{d4flux-new}
\hat F = \left(\begin{array}{ccccc}
0 &  f & 0&0&0\\
-  f & 0 &  g &0 & 0\\
0 & - g & 0 &0 &0\\
0&0&0&0&0\\
0&0&0&0&0\end{array} \right),
\ee  
while the D2 carries no flux at all, lying along 1, 2 directions.  For this system, following \cite{Jia:2019hbr}, we have the open string one-loop annulus amplitude as
\be\label{42annulusA}
\Gamma_{4, 2} = \frac{2 V_{3} \sqrt{g^{2} - f^{2}}}{(8 \pi^{2} \alpha')^{3/2}} \int_{0}^{\infty}  \frac{d t}{t^{3/2}} \, e^{- \frac{y^{2} t}{2\pi \alpha'}} \frac{\left(\cosh\frac{\pi t}{2} - \cosh\pi \bar\nu t\right)^{2}}{\sinh \frac{\pi t}{2} \sinh\pi \bar\nu t} \prod_{n = 1}^{\infty} \bar Z_{n},
\ee 
where $\bar Z_{n}$ is the same as the $Z_{n}$ given in (\ref{zn}) if we replace the $\nu$ parameter there by the present $\bar \nu$.  Here $\bar\nu$ continues to be given by (\ref{parameternup}). Again, if we make the identifications of the fluxes as given in (\ref{parameterident}),  we then have $\bar \nu = \nu$  (therefore $\bar Z_{n} = Z_{n}$) and apart from a trivial factor $2 g V_{2}/(8 \pi^{2} \alpha')$, the above amplitude is identical to the D4/D0 one given in (\ref{40annulusA}).  Here we have denoted $V_{3} = V_{1} V_{2}$.  

Just as for the D2/D2 system discussed in the previous section, the present D4/D2 amplitude (\ref{42annulusA}) is also an invariant under the D2 worldvolume Lorentz transformation. So we can repeat what we did in the previous section to give an explanation to each of the three subcases discussed above for the D4/D0.  Since the detail here is different from that for the D2/D2 or D2/D0 system discussed in the previous section, we give each of the subcases a discussion in order below.

For the subcase of $g^{2} > f^{2}$, corresponding to $0 < {\hat f}^{2} < 1$, we can use the following D2 worldvolume Lorentz transformation 
\be\label{42LT}
(\Lambda_{4, 2})_{\alpha}\,^{\beta} = \left(\begin{array}{cc}
(\Lambda_{2, 2})_{\alpha'}\,^{\beta'} & 0\\
0 & I_{2 \times 2} \end{array}\right),
\ee
to transform the flux on the D4 in the D4/D2 system to a pure magnetic flux on the D4 given below 
\be\label{42transf-f}
\bar F_{\alpha\beta} = \left(\begin{array}{cc}
\bar F_{\alpha'\beta'} & 0\\
 0 & 0_{2\times 2}
  \end{array}\right),
 \ee
where $\alpha, \beta =0, 1, 2, 3, 4$; $\alpha', \beta' = 0, 1, 2$ and $(\Lambda_{2, 2})_{\alpha'}\,^{\beta'}$ is just the one given in (\ref{g-Lorentz}) and the flux $\bar F_{\alpha'\beta'}$ is the one given in (\ref{transf}).   In the above, $I_{n \times n}$ stands for a  $n \times n$ unit matrix while $0_{n \times n}$ stands for a $n \times n$ zero matrix.  The resulting system is one consisting of a D4 carrying a pure magnetic flux along 1, 2 directions, while the D2 remains carrying no flux. Since the pure magnetic flux along 1, 2 directions within the D4 stands for D2 branes along 3, 4 directions but delocalized along 1, 2 directions. In other words, we have a non-threshold bound state (D2, D4) with the D2 branes along 3, 4 directions while delocalized along 1, 2 directions.  The delocalized  D2 branes have no interaction with the D2 in the system which can be easily seen via T-dualities along 1, 2 directions.  By the T-dualities, the delocalized D2 become D4 while the  original D2 becomes D0 and we know that there is no interaction between the D4 and the D0 since their dimensionality differs by 4.   So the interaction between the transformed (D2, D4) and the D2 is purely between the D4 and the D2 which is known to be attractive. So this gives an explanation to the attractive interaction between the original D4 carrying the given flux (\ref{d4flux-new}) and the D2 of the original D4/D2 system. This in turn gives an explanation to the attractive interaction between the D4 and the D0 for this subcase.

For the subcase $g^{2} = f^{2}$, corresponding to the ${\hat f}^{2} = 1$, this D4/D2 system is equivalent to a D4/D2 one with both D4 and D2 carrying no flux at all via the Lorentz transformation given in (\ref{g-Lorentz}) by taking the limit $ f^{2} \to  g^{2}$.  The resulting flux given in (\ref{transf}) simply vanishes in this limit.  Unlike the D2/D0 or D2/D2, the present amplitude as given in (\ref{40annulusA-new}) or from (\ref{42annulusA}) by taking this limit does not vanish and turns out to be  positive (therefore the interaction is attractive).  Now this can easily be understood. Since the amplitude for the original D4/D2 is equivalent to the Lorentz transformed one for which neither D4 or D2 carries any flux. It is known that the interaction between a simple D4 carrying no flux and a simple D2 carrying no flux either, placed parallel at a separation, is attractive and is precisely given by the limit taken (See  \cite{Jia:2019hbr} for example). 

For the subcase $g^{2} < f^{2}$, corresponding to the $1 < {\hat f}^{2} < 1 + {\hat g}^{2}$, we can now use the Lorentz transformation given in (\ref{42LT}) but now with $(\Lambda_{2, 2})_{\alpha'}\,^{\beta'}$ given by (\ref{f-Lorentz}) to transform the flux on D4 to a pure electric  one given in (\ref{42transf-f}) but now with the $\bar F_{\alpha'\beta'}$  given by (\ref{e-transf}).  The resulting system is one consisting of
D4 carrying a pure electric flux and the D2 carrying no flux.  The system gives rise to the open string pair production as shown in \cite{Jia:2019hbr} which says that the original D4/D2 gives rise to the open string pair production.  This in turn says that the D4/D0 system gives rise to the open string pair production in this subcase even though the underlying mechanism is different as discussed earlier. 

\section{The D6/D0 case}
We move to the final case of this type.  The flux on D6 is given by (\ref{dpflux}) for $p = 6$. Again following \cite{Jia:2019hbr},  we compute the open string one-loop annulus amplitude as
\be\label{60annulusA}
\Gamma_{6, 0} = - \frac{V_{1} \sqrt{1 - {\hat f}^{2}}}{(8 \pi^{2} \alpha')^{1/2}} \int_{0}^{\infty} \frac{d t}{t^{3/2}} \, e^{- \frac{y^{2} t}{2\pi \alpha'}} \frac{\left(\cosh^{2} \frac{\pi t}{2} - \cosh^{2}\frac{\pi \nu t}{2}\right) \tanh \frac{\pi \nu t}{2} }{\sinh^{2} \frac{\pi t}{2}} \prod_{n = 1}^{\infty} Z_{n},
\ee
where $y$ is the separation between the D6 and D0, the parameter $\nu$ is again given by (\ref{parameternu}) and
\be\label{60zn}
Z_{n} = \frac{\left(1 - 2 |z|^{2n} \cosh \pi \nu t + |z|^{4n}\right)^{2} \prod_{j = 1}^{2}\left(1 - 2 |z|^{2n} \cosh \pi (\nu + (-)^{j}) t + |z|^{4n}\right)}{\left(1 - |z|^{2n}\right)^{2} \left(1 - 2 |z|^{2n} \cosh\pi t + |z|^{4n}\right)^{2} \left(1 -  2 |z|^{2n} \cosh 2\pi \nu t + |z|^{4n}\right)},
\ee
with $|z| = e^{- \pi t}$.  As before, we have three subcases to consider and we will discuss each of them below in order.\\
\noindent
{\bf Subcase 1: $0 < {\hat f}^{2} < 1$.}\\
For this subcase, we again have $\nu \in (0, 1/2)$ when $\hat g \neq 0$ from (\ref{parameternu}).  Then it is clear that the amplitude (\ref{60annulusA}) is negative and so the interaction is repulsive.  We will explain this later on. The integrand of this amplitude has no singularities, in particular no simple poles, for any finite $t$ in the allowed range of ${\hat f}^{2}$. So there is no open string pair production. For the present case, there is no potential blowup behavior even for  large $t$.  Note that for large $t$, the integrand behaves like, noticing $Z_{n} \to 1$, 
\be
\sim  \frac{e^{- \frac{y^{2} t }{2 \pi \alpha'}}}{t ^{3/2}} \to 0,
\ee
therefore there is no tachyonic instability to occur for any $y$.  This is a typical behavior when the interaction is repulsive as noticed also in \cite{Jia:2019hbr}.\\
\noindent
{\bf Subcase 2: ${\hat f}^{2} = 1$.}\\
 For this subcase,  we have $\nu = 0$ from (\ref{parameternu}) and $Z_{n} = 1$ from  (\ref{60zn}). It is clear that the amplitude (\ref{60annulusA})  vanishes but the reason appears different from the corresponding one for the D2/D0 system discussed in section 2. We will come to discuss this later on. \\
 \noindent
 {\bf Subcase 3: $1 < {\hat f}^{2} < 1 + {\hat g}^{2}$.}\\
  For this subcase, as before, the parameter $\nu$ from (\ref{parameternu}) becomes imaginary and we set $\nu = i \nu_{0}$ with $\nu_{0} \in (0, \infty)$.  The amplitude (\ref{60annulusA}) becomes
  \be\label{60annulusAmp}
  \Gamma_{6, 0} = \frac{V_{1} \sqrt{{\hat f}^{2} - 1}}{(8 \pi^{2} \alpha')^{1/2}} \int_{0}^{\infty} \frac{d t}{t^{3/2}} \, e^{- \frac{y^{2} t}{2\pi \alpha'}} \frac{\left(\cosh^{2} \frac{\pi t}{2} - \cos^{2}\frac{\pi \nu_{0} t}{2}\right) \tan \frac{\pi \nu_{0} t}{2} }{\sinh^{2} \frac{\pi t}{2}} \prod_{n = 1}^{\infty} Z_{n},
\ee
where 
\be\label{60zn-new}
Z_{n} = \frac{\left(1 - 2 |z|^{2n} \cos \pi \nu_{0} t + |z|^{4n}\right)^{2} \prod_{j = 1}^{2}\left(1 - 2 |z|^{2n} \cosh \pi (i \nu_{0} + (-)^{j}) t + |z|^{4n}\right)}{\left(1 - |z|^{2n}\right)^{2} \left(1 - 2 |z|^{2n} \cosh\pi t + |z|^{4n}\right)^{2} \left(1 -  2 |z|^{2n} \cos 2\pi \nu_{0} t + |z|^{4n}\right)},
\ee
with $|z| = e^{- \pi t}$.  The $\nu_{0}$ is given by  (\ref{parameternu0}).  It is clear that the integrand of the amplitude (\ref{60annulusAmp}) has simple poles along the $t$-axis occurring at
\be
\frac{\pi \nu_{0} t_{k}}{2} = \frac{\pi (2 k - 1)}{2} \to t_{k} = \frac{2 k - 1}{\nu_{0}}, \quad k = 1, 2, \cdots,
\ee
indicating the system decays via the open string pair production with the decay rate 
\bea\label{60decayr}
{\cal W}_{6, 0}  & =& - \frac{2 {\rm Im} \Gamma_{6, 0}}{ V_{1}}\nn
&=& \frac{2^{2} \sqrt{{\hat f}^{2} - 1}}{\nu_{0} (8\pi^{2}\alpha')^{1/2}} \sum_{k = 1}^{\infty} \left(\frac{\nu_{0}} {2 k - 1}\right)^{3/2} \, e^{- \frac{(2 k - 1) y^{2}}{2\pi \nu_{0}\alpha'}} \frac{\cosh^{2} \frac{(2 k - 1) \pi}{2 \nu_{0}}}{\sinh^{2} \frac{ (2 k - 1) \pi}{2 \nu_{0}}} \nn
&\,& \times\prod_{n =1}^{\infty} \frac{\left(1 + |z_{k}|^{2n}\right)^{4} \left(1 + 2 |z_{k}|^{2n} \cosh \frac{(2 k - 1) \pi}{\nu_{0}} + |z_{k}|^{4n}\right)^{2}}{\left(1 - |z_{k}|^{2n}\right)^{4} \left(1 - 2 |z_{k}|^{2n} \cosh \frac{(2 k - 1)\pi}{\nu_{0}} + |z_{k}|^{4n}\right)^{2}},
\eea
where $|z_{k}| = e^{- (2 k - 1)\pi/\nu_{0}}$.  The open string pair production rate is given by the $k = 1$ term of the above decay rate, following \cite{nikishov},  as
\be\label{60opr}
{\cal W}^{(1)}_{6, 0} =  \frac{2^{2}  \nu_{0}^{1/2} \sqrt{{\hat f}^{2} - 1}}{(8\pi^{2}\alpha')^{1/2}} \, e^{- \frac{y^{2}}{2\pi \nu_{0}\alpha'}} \frac{\cosh^{2} \frac{\pi}{2 \nu_{0}}}{\sinh^{2} \frac{\pi}{2 \nu_{0}}} 
\prod_{n = 1}^{\infty} \frac{\left(1 + |z_{1}|^{2n}\right)^{4} \left(1 + 2 |z_{1}|^{2n} \cosh \frac{ \pi}{\nu_{0}} + |z_{1}|^{4n}\right)^{2}}{\left(1 - |z_{1}|^{2n}\right)^{4} \left(1 - 2 |z_{1}|^{2n} \cosh \frac{\pi}{\nu_{0}} + |z_{1}|^{4n}\right)^{2}},
\ee
where $|z_{1}| = e^{- \pi/\nu_{0}}$.  The mechanism of this open pair production is the same as that for D2/D0, i.e., by braking the virtual open string and the anti open string pair connecting the D6 and the D0. We will not repeat this same discussion here.

We now use the D6/D2 system to give an explanation to each of the above three subcases. The flux on this D6 is again structurally the same as that given in (\ref{dpflux}) for p = 6 for the D6 in the D6/D0 but we denote it differently as before, i.e., the electric flux or the magnetic one without a hat.  The D2 carries no flux at all, lying along 1, 2 directions.  We then follow  \cite{Jia:2019hbr} to compute the open string one-loop annulus amplitude with the brane separation $y$, taken the same as that for D6/D0, as
\be\label{62annulusA}
\Gamma_{6, 2} = - \frac{2 V_{3} \sqrt{g^{2} - f^{2}}}{(8 \pi^{2} \alpha')^{3/2}} \int_{0}^{\infty} \frac{d t}{t^{3/2}} \frac{\left(\cosh^{2} \frac{\pi t}{2} - \cosh^{2} \frac{\pi \bar\nu t}{2}\right) \tanh\frac{ \pi \bar\nu t}{2}} {\sinh^{2} \frac{\pi t}{2} } \prod_{n = 1}^{\infty} \bar Z_{n},
\ee
where the parameter $\bar \nu$ is again determined by (\ref{parameternup}) and $\bar Z_{n}$ is identical to $Z_{n}$  given in (\ref{60zn}) if we replace the $\nu$ there by the present $\bar\nu$. With these, again up to a trivial factor of $2 g V_{2}/(8 \pi^{2}\alpha')$, the present amplitude (\ref{62annulusA}) is identical to the one (\ref{60annulusA}) for the D6/D0 if we also make the flux identifications of (\ref{parameterident}).  For this reason, we can use the amplitude (\ref{62annulusA}), which is invariant under the D2 worldvolume Lorentz transformation, to give an alternative understanding of the underlying physics for each subcase discussed above for the D6/D0 system, following the same steps as given for the D2/D2 case discussed in section 2.

The key is that the amplitude (\ref{62annulusA}) for the D6/D2 is physically equivalent to the D2 worldvolume Lorentz transformed one.  For $g^{2} > f^{2}$, i.e., $0 < {\hat f}^{2} < 1$, we can use the following Lorentz transformation  
\be\label{62LT}
(\Lambda_{6, 2})_{\alpha}\,^{\beta} = \left(\begin{array}{cc}
(\Lambda_{2, 2})_{\alpha'}\,^{\beta'} & 0\\
0 & I_{4 \times 4} \end{array}\right),
\ee
to transform the original flux on the D6 to the one with a pure magnetic flux as 
\be\label{transf-f}
\bar F_{\alpha\beta} = \left(\begin{array}{cc}
\bar F_{\alpha'\beta'} & 0\\
 0 & 0_{4\times 4}
  \end{array}\right),
 \ee
 where  $\alpha, \beta =0, 1, \cdots 6$; $\alpha', \beta' = 0, 1, 2$; $(\Lambda_{2, 2})_{\alpha'}\,^{\beta'}$ is just the one given in (\ref{g-Lorentz}) and $\bar F_{\alpha'\beta'}$ is given by (\ref{transf}) as a pure magnetic flux along 1, 2 directions. This magnetic flux  represents D4 branes lying along 3, 4, 5, 6-directions but delocalized along 1, 2 directions within the D6 brane.  So the transformed system consists of a non-threshold bound state (D4, D6) and a D2.
 
 To make it easy to understand the interaction between the (D4, D6) and the D2, we T-dualize along 1, 2 directions, respectively.  We have then (D4, D6) $\to$ (D6, D4) for which the D4 branes in the (D4, D6) become D6 while the D6 becomes D4. At the same time the D2 becomes D0. So we end up with a system of (D6, D4) and D0.  It is known that there is no interaction between the D4 in (D6, D4) and the D0 since their dimensionality differs by 4 while the interaction between the D6 and D0 is repulsive (as shown in \cite{Jia:2019hbr}  for example). So the net interaction between the (D6, D4) and the D0 is repulsive.  This implies that the interaction between the (D4, D6) and D2 of the transformed system is repulsive and so is the interaction between the D6 and the D2 of the original system D6/D2 before the Lorentz transformation.  This gives an explanation of this subcase.
 
 For the subcase of $g^{2} = f^{2}$, corresponding to ${\hat f}^{2} = 1$, the original D6/D2 system after the transformation becomes one with neither the D6 nor the D2 carries any flux.  Since there is no interaction between  a  D6 and a D2 (their dimensionality differs by 4),  placed parallel at separation, when neither of them carries any flux, this gives an explanation to the zero net interaction between the D6 and the D2 before the Lorentz transformation, therefore to the D6/D0 system for the present subcase. 
 
 For the subcase of $g^{2} < f^{2}$, corresponding to $1 < {\hat f}^{2} < 1 + {\hat g}^{2}$, the flux on D6 of the original D6/D2 system can be transformed to  a pure electric one on the D6 by a Lorentz transformation similar to (\ref{62LT}) but now with the  $(\Lambda_{2, 2})_{\alpha'}\,^{\beta'}$ being given by (\ref{f-Lorentz}) and the transformed flux similar to (\ref{transf-f}) with the $\bar F_{\alpha'\beta'}$  being given by (\ref{e-transf}).  So the transformed D6/D2 is the one with the D6 carrying a pure electric flux along 1 direction while the D2 remains as before. Our previous studies  \cite{Jia:2019hbr} say that there must exist the open string pair production and so is the original D6/D2. This in turn says that there exists the open string pair production for the D6/D0 system even though the mechanism is different as discussed earlier. 
 
 \section{Conclusion}
 In this note, we explore the condition under which a finite non-vanishing open string pair production can occur for the system of Dp/D0 with $p = 2, 4, 6$, respectively.  It turns out that a pure electric flux on the Dp is insufficient for this purpose while keeping this brane stable  though this is enough when the other brane has its spatial dimensionality greater than zero.  A non-vanishing magnetic flux has to appear in a way  on the Dp such that it  allows the absolute value of the dimensionless electric flux to be greater than unity which is needed for the pair production to occur while still maintaining the Dp stable (for example, a uniform choice for this is given in the previous sections). In addition to this subcase for each given $p > 0$, there exist the other two subcases, corresponding to $ 0 <  {\hat f}^{2} < 1$ and ${\hat f}^{2} = 1$, for which we give a physical understanding why the underlying interaction between the Dp and D0 is the way it is supposed to be. For this, we replace the system Dp/D0 by an essentially equivalent Dp/D2, so long the interaction is concerned, when the electric flux and the magnetic one on the Dp in Dp/D2 (noting that the D2 carries no flux) are related to their correspondences on the Dp in Dp/D0 in a specific way as given in (\ref{parameterident}).  The interaction nature of this Dp/D2 can be understood via its Lorentz transformed one which is simple enough to be understood easily.  This can in turn be used to understand the interaction nature of Dp/D0 system including the open string pair production though the mechanism for this is different.


\section*{Acknowledgments}
The author acknowledges the support by grants from the NNSF of China with Grant No: 12275264 and 12247103.

\section*{A  The pair production rate vs the vacuum decay rate \label{A}}
In the case of a constant electric field applied to the 4-dimensional  QED vacuum, the vacuum decay rate  gives rise to the pair production probability per unit volume and time, called Schwinger Mechanism \cite{Schwinger:1951nm}, and is given by
\be\label{drate}
{\cal W} = \frac{2 (e E)^{2}}{(2 \pi)^{3}} \sum_{n = 1}^{\infty} \frac{1}{n^{2}} e^{- \frac{n \pi m^{2}}{e E}},
\ee
from the so-called vacuum persistence probability
\be
P = |\langle 0, {\rm out}|0, {\rm in}\rangle|^{2} = e^{- \int d^{4} x\, {\cal W}}.
\ee
This formula has been derived in various ways and its generalization to the case of inhomogeneous  background field has also been considered, for examples, see \cite{Brezin:1970xf, Kim:2007pm} .  In essence, this ${\cal W}$ counts how the original vacuum decays via the pair production under the action of the applied field via the pair production, but does not count the subsequent fate of the pairs so produced per unit volume and time (for example, the produced pair can annihilate afterwards). So long the detection is concerned, the average pairs produced per unit volume and time  are more relevant and this gives the so-called pair production rate.  Nikishov noticed this difference \cite{nikishov} and thought that the pair production rate should be computed from the particle mean number $\langle 0, {\rm in}| a^{+}_{\bf p} a_{\bf p } |0, {\rm in}\rangle$ or its anti particle one (the two are the same). As Nikishov showed in \cite{nikishov}, the total particle or anti-particle mean number from all possible modes per unit volume and time gives just the first term, i.e., the $n = 1$ term,  in the above (\ref{drate}).  Note that the computation of this rate is  independent of the one in deriving the pair production probability per unit volume and time ${\cal W}$ (\ref{drate}).  This has been clearly discussed in \cite{Tanji:2008ku} and the case including the  collinear constant electric and magnetic fields and for different spins has also been discussed in \cite{Kim:2003qp, Kruglov:2004jx, Gavrilov:2006jb}.

 Let us give here a simple illustration of this for a massive charged scalar with mass $m$. The distribution function of produced charged particles $n_{\bf p}$ is same as that of anti-charged ones under the action of the constant electric field $E$ applied to the vacuum and both are given  in equation (35) in \cite{Tanji:2008ku} as
\be\label{df}
n_{\bf p} = \bar n_{\bf p} = e^{- \frac{\pi (m^{2} + p^{2}_{x} + p^{2}_{y})}{e E}},
\ee
where $p_{x}$ and $p_{y}$ are the momenta perpendicular to the direction of the applied electric field $E$ (assuming $E$ along the $z$-direction).  For this case, the decay rate of the vacuum or the pair production probability per unit volume and time is the standard one and is derived, for example, in equation (40) in \cite{Tanji:2008ku} as
\be\label{scalar-dr}
{\cal W} = \frac{( e E)^{2}}{(2\pi)^{3}} \sum_{n = 1}^{\infty} \frac{(- 1)^{n + 1}}{n^{2}} \, e^{- \frac{\pi m^{2}}{e E}}.
\ee 
Following \cite{nikishov}, we can obtain the pair production rate by first computing the average charged particle number $N$ (or the average anti-charged particle number $\bar N$) by the integration of $n_{\bf p}$ in the phase space as
\bea
N &=& \bar N = \frac{V}{(2\pi)^{3}} \int d p_{x} d p_{y} d p_{z} n_{\bf p} = \frac{V}{(2\pi)^{3}} \int d p_{x} d p_{y} d p_{z}  e^{- \frac{\pi (m^{2} + p^{2}_{x} + p^{2}_{y})}{e E}}\nn
&=& \frac{V T (e E)^{2}}{(2\pi)^{3}} e^{- \frac{\pi m^{2}}{e E}},
\eea
where we have used $p_{z} = - e E t$ when the charged particle/anti-particle so produced in the pair production occurs at rest such that $\int_{- \infty}^{\infty} d p_{z} = e E \int_{- \infty}^{\infty} d t =  e E T$ with $T$ denoting the total time interval (In other words, the momentum $p_{z}$ for the produced charged particle/anti-particle is due to the action of the applied electric field $E$).  This gives immediately the particles or the anti-particles produced per unit volume and time, i.e., the pair production rate, as
\be
\frac{N}{VT} = \frac{\bar N}{VT} = \frac{(e E)^{2}}{(2\pi)^{3}} \, e^{- \frac{\pi m^{2}}{e E}},
\ee
which is nothing but the first term in the decay rate given in (\ref{scalar-dr}).

Another interesting way of interpreting this difference \cite{Cohen:2008wz} is that in deriving the pair production probability per unit volume and time the correlation between pair productions at different spacetime points is not counted, which is consistent with the picture given above for ${\cal W}$, while for the pair production rate, this correlation in coordinate spacetime should be considered since when the pair is produced, the momentum modes are viewed independently and as such they are correlated in coordinate spacetime.  With this consideration, the authors in \cite{Cohen:2008wz} indeed showed this difference and found the agreement with Nikishov's result \cite{nikishov}. Other related discussions are referred to  \cite{ Nikishov:2003ig, Kim:2003qp, Kruglov:2004jx, Kim:2007pm, Gavrilov:2006jb}.

\end{document}